\documentclass[twocolumn]{article}
\usepackage[paper=letterpaper, dvips, top=1.5cm, left=1.5cm, right=1.5cm, foot=1.5cm, bottom=3cm]{geometry}

\usepackage{setspace}
\usepackage[paper=letterpaper]{geometry}
\usepackage{graphicx,subfigure}
\usepackage{cite,url}
\usepackage{amsmath} 
\usepackage{amssymb}  
\usepackage{xcolor}

\usepackage{breqn}
\newtheorem{theorem}{Theorem}
\newtheorem{definition}{Definition}

\renewcommand\appendix{\par
  \setcounter{section}{0}
  \setcounter{subsection}{0}
  \setcounter{figure}{0}
  \setcounter{table}{0}
  \renewcommand\thesection{Appendix \Alph{section}}
  \renewcommand\thefigure{\Alph{section}\arabic{figure}}
  \renewcommand\thetable{\Alph{section}\arabic{table}}
}
    
\begin{document}
\title{Describing Function-based Approximations of Biomolecular Systems}
\author{
        Abhishek Dey$^{1}$, Shaunak Sen$^{2}$\\
        Department of Electrical Engineering\\
        Indian Institute of Technology Delhi\\
        Hauz Khas, New Delhi 110016, INDIA\\
        E-mail: abhishek.dey@ee.iitd.ac.in$^{1}$, shaunak.sen@ee.iitd.ac.in$^{2}$
}

\date{{\color{blue}Accepted in IET Systems Biology, doi: 10.1049/iet-syb.2017.0026}}
\twocolumn[
  \begin{@twocolumnfalse}
\maketitle
\noindent
\begin{abstract} 
Mathematical methods provide useful framework for the analysis and design of complex systems.
In newer contexts such as biology, however, there is a need to both adapt existing methods as well as to develop new ones.
Using a combination of analytical and computational approaches, we adapt and develop the method of describing functions to represent the input-output responses of biomolecular signalling systems.
We approximate representative systems exhibiting various saturating and hysteretic dynamics in a way that is better than the standard linearization.
Further, we develop analytical upper bounds for the computational error estimates.
Finally, we use these error estimates to augment the limit cycle analysis with a simple and quick way to bound the predicted oscillation amplitude. 
These results provide system approximations that can add more insight into the local behaviour of these systems than standard linearization, compute responses to other periodic inputs, and to analyze limit cycles.
\end{abstract}
\end{@twocolumnfalse}
]

\section{Introduction}\label{sec1}

Design of biomolecular systems can enable applications in agriculture, medicine and manufacturing~\cite{c1}.
Complementarily, analyzing how naturally occuring biomolecular interactions determine cellular behaviour is a fundamental problem in biology~\cite{c2}.
Mathematical frameworks are useful for both these objectives.
These provide system representations to test and compare different design choices as a guide to the actual implementation.
These also help to develop useful insight into how system interactions can combine to generate the overall behaviour. Mathematical models used in these cases are typically complex, both due to their large scale as well as the inherent nonlinearity, making their analysis challenging.
Therefore, there is a need to adapt existing mathematical methods as well as develop these and new ones for the study of such problems.

Biomolecular systems are frequently represented as ordinary differential equations, formulated based on the principles of mass action.
The variables in these equations are the concentrations of various biomolecular species that evolve in time depending on their interactions with each other.
One approach to study these mathematical models is exhaustive numerical computations, which can catalog all possible system behavior, complemented with simpler calculations to understand the key underlying principles~\cite{c3,chemotaxis}.
Another approach is theoretical, exploiting the inherent structure to infer system behavior. 
An example is the theory of monotone systems~\cite{c4}.
Intermediary approaches may also exist using various approximations to understand system behaviour.
Indeed, methods of linear systems theory have often been used to characterize biomolecular system behaviour, such as impulse responses in bacterial chemotaxis~\cite{block1982impulse} as well as frequency responses in the osmolarity pathway of the yeast \emph{Saccharomyces cerevisiae} mediated by the MAP Kinase cascade~\cite{c11, freq_osmo_adaptation, fluoro_microscopy} and in the Galactose metabolic pathway, also in the yeast \emph{Saccharomyces cerevisiae}~\cite{bennett2008metabolic}. Another example of such an intermediary approach is the describing function technique~\cite{c5,c7}, where the frequency response of a nonlinear system to a sinusoidal input of a particular amplitude is approximated by the first harmonic of the resulting output response. 
This is widely used in classical control engineering to estimate limit cycle behavior as well as to replace nonlinear input-output responses with corresponding linear approximation.
These linear approximations can then be analyzed using the well developed tools of linear systems theory. 
In fact, this technique has been applied to analyze biomolecular oscillations~\cite{goodwin} and to approximate input-output maps in biomedical contexts~\cite{reviewer_ref}.
These results present important early work in using this technique for biological systems.

There are at least three striking aspects related to using a describing function-based linearization to approximate the input-output response of a biomolecular signalling system.
One, describing functions naturally allow the analysis of finite amplitude inputs, as opposed to infinitesimal amplitude inputs in the standard linearization.
These may be more relevant to actual biomolecular contexts, such as in the experimental studies of frequency response mentioned above~\cite{c11, freq_osmo_adaptation, fluoro_microscopy,bennett2008metabolic} and provide additional insight into the system behaviour.
Two, nonlinearities typically analyzed using describing functions in the classical contexts are static nonlinearities, like saturation or hysteresis. Contrastingly, the same nonlinearities in biomolecular contexts can have a dynamic character because of the underlying nonlinear biomolecular interactions are embedded in the overall system dynamics.
Three, there may be error involved in the approximation that may depend both on input frequency as well as inherent system parameters, which may be important to quantify.
Given these, the describing function-based approximation of these input-output responses, including the dependence on system parameters, and the nature of approximation error is generally unclear.

Here, we aim to approximate these systems and estimate the resultant error.
For this, we used the technique of describing functions, analytically, where possible, as well as computationally.
We computed approximations for representative systems with input-output responses exhibiting saturation dynamics with different slopes as well as hysteretic dynamics (formed the basis of preliminary investigation, \cite{deysen}).
Next, we computed the approximation error and developed a theoretical error bound for these kind of systems.
Finally, we used these error estimates to augment the classical describing function-based limit cycle analysis by providing a simple way to estimate the range of oscillation amplitudes.
These results adapt the existing method of describing functions for the study of biomolecular systems and should be useful both in analysis and design.
\section{Computation of Describing Function Approximation}
\label{sec2}

\noindent \emph{Describing Function Method}

Consider a nonlinear system governed by equations,
 \begin{eqnarray}
 \dot{\mathbf{x}} &=& \mathbf{f}(\mathbf{x},u), \nonumber \\
 y &=& \mathbf{C}\mathbf{x},
 \label{eq:Nonlinear System}
 \end{eqnarray}
where, $\mathbf{x}$ is a vector of species concentrations, $u$ is input which can be a system parameter, $\mathbf{C}$ is a row vector and $y$ is output and typically, $\mathbf{f}$ is a nonlinear map. 
We aim to approximate the input-output system from $u$ to $y$ using describing function~\cite{c5}. To compute the approximation we set the input to be, $u = u_{0} + b\sin(\omega t)$, where $b$ is the forcing amplitude, $\omega$ is the forcing frequency and $u_{0}$ is the input bias. The describing function approximation $G(j\omega, b, u_{0})$ can be obtained by taking the first harmonic of the output in response to sinusoidal input. This is defined as,
\begin{eqnarray}
Re\{G(j\omega, b, u_{0})\} &=& \dfrac{\omega}{\pi b}\int_0^{2\pi}y(t)\sin(\omega t)dt, \nonumber \\
Im\{G(j\omega, b, u_{0})\} &=& \dfrac{\omega}{\pi b}\int_0^{2\pi}y(t)\cos(\omega t)dt.
\label{eq:ComputationalDF}
\end{eqnarray}
The magnitude and phase of the approximation can be calculated as,
\begin{eqnarray}
|G| &=& \sqrt{Re\{G\}^2 + Im\{G\}^2}, \nonumber \\
\angle G &=& \tan^{-1} \frac{Im\{G\}}{Re\{G\}}
\end{eqnarray}
respectively. 

Next, we use this technique to approximate canonical biomolecular signalling systems. Here, Eq. \ref{eq:Nonlinear System} is an ordinary differential equation model of the system under consideration obtained using the law of mass action. As, biomolecular signals can not be negative, we use the constraint $b<u_{0}$, so that the input is always positive.\\
\noindent \emph{Example 1: Biomolecular System with Saturating Input-Output Map}

\begin{figure}[!htb]
\center
\includegraphics[scale=0.43]{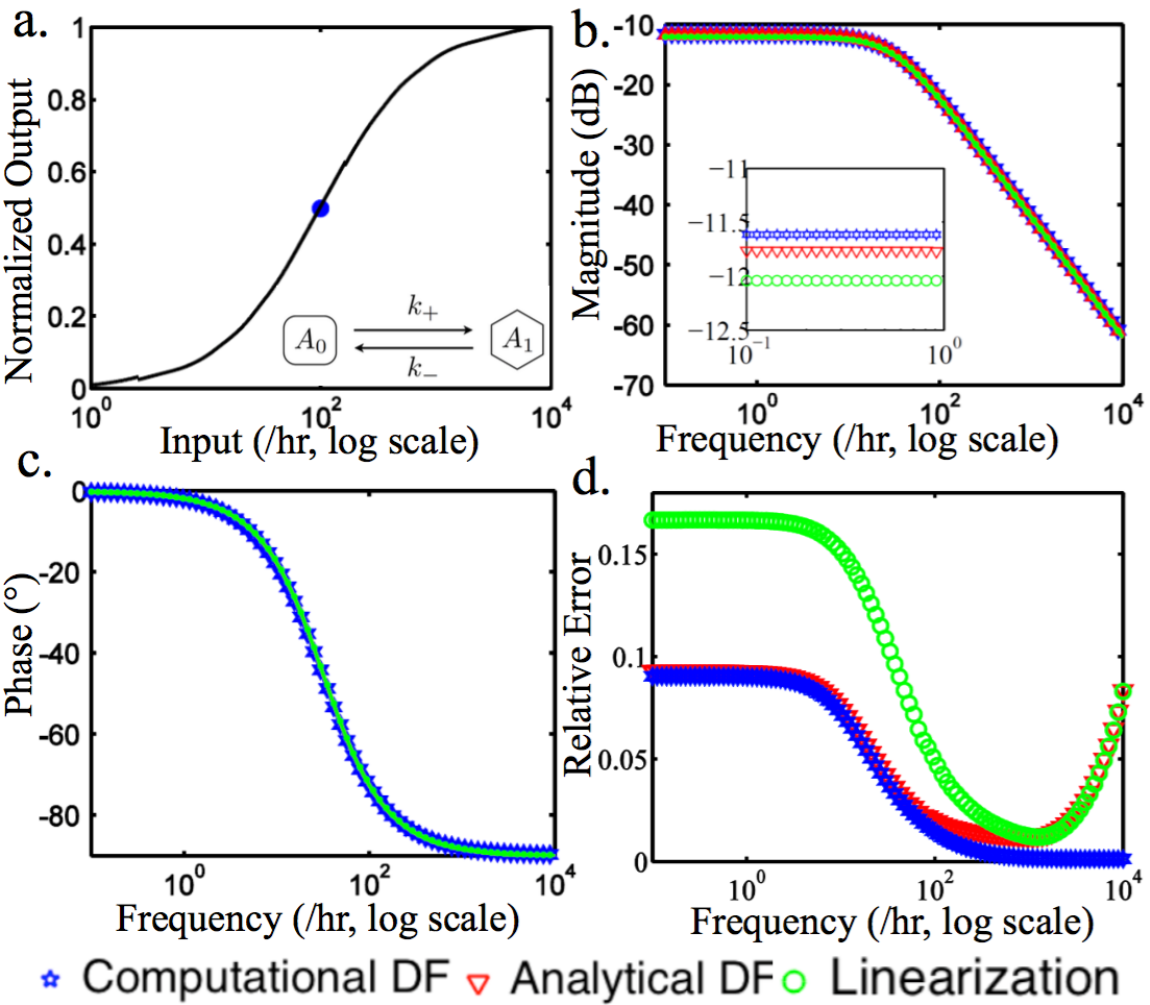}
\caption{Approximation of the simple biomolecular system and error estimate. a. Black line is the steady-state input-output response for parameters $A_T = 100nM$, $k_- = 100/hr$, $k_{+0} = k_-$, $b = k_{+0}/2$. Blue dot is the operating point used for computation of linearization. Schematic of biomolecular system is shown in the inset.
b. Red triangle shaped, blue pentagon shaped and green circle shaped markers are the magnitude plots of analytical, computational describing function-based approximations and direct linearization respectively for same parameter set, frequency is varied logarithmically in the range $0.1/hr$--$10^4/hr$. c. Phase plots corresponding to b. d. Corresponding error plots.}
\label{fig:figure1}
\end{figure}

One of the simplest signal transduction mechanisms that is representative of diverse signaling contexts is of a biomolecular species that can exist in two states~\cite{c3, retina, stock}. These can interconvert among each other (Fig. \ref{fig:figure1}a, inset) depending on the input level. Typically, one of these states has biological activity and can serve as the output.
Consider a simple model of this with a biomolecular species ($A$) that can interconvert between two forms ($A_0$ and $A_1$) at certain forward and reverse rates ($k_+$ and $k_-$).
In this two-state model, the input, such as temperature or pheromone levels, can be modeled as modulating the rate $k_+$ and the output as the concentration of $A_1$, the active form of the protein $A$. A mathematical model in the context of biomolecular signal transduction can be obtained using mass action kinetics,
\begin{eqnarray}
\dfrac{dA_1}{dt} = k_+(A_T-A_1) - k_-A_1.
\label{eq:eqn1}
\end{eqnarray}
Here, the total concentration of the protein is $A_T$ ($=A_0 + A_1$), which is constant in the simplest scenario considered here.
Due to the presence of the term $k_+A_1$, where the input term and the output multiply each other, this is a nonlinear equation. The input-output response at steady-state can be obtained by setting $\dfrac{dA_1}{dt} = 0 \implies A_1 = A_T \dfrac{k_+}{k_+ + k_-}$ (Fig. \ref{fig:figure1}a).

To compute the approximation of the overall response, we set the input to be $k_+ = k_{+0} + b\sin(\omega t)$ and describing function approximation is computed using the output obtained numerically from MATLAB ode23s solver in response to the above input from Eqn. \ref{eq:ComputationalDF}. The results of this computation are shown in Fig. \ref{fig:figure1}b, c.

To obtain an analytical approximation, we set $A_1 = A_{10} + A_{1b}\sin(\omega t +\theta)$~\cite{c5} in Eq. (\ref{eq:eqn1}) and collect the like terms to obtain expressions for $A_{10}$, $A_{1b}$, and $\theta$, (the detailed steps are in Supplementary Material S1)
 \begin{eqnarray}
A_{10} &=& A_T \dfrac{k_{+0} - \alpha}{k_- + k_{+0} - \alpha}, \alpha =  \dfrac{1}{2}b^2\dfrac{k_{+0} + k_-}{\omega^2 + (k_{+0} + k_-)^2},\nonumber \\
A_{1b} &=& (A_T - A_{10})\dfrac{b}{\sqrt{\omega^2 + (k_{+0} + k_-)^2}} ,\nonumber \\
\theta &=& -\tan^{-1}\big( \dfrac{\omega}{k_{+0} + k_-} \big). 
\label{eq:eqn2}
\end{eqnarray}
The describing function approximation is (Fig. \ref{fig:figure1}b, c),
\begin{eqnarray}
G(j\omega, b, k_{+0}) &=& \dfrac{A_{1b}}{b}e^{j(\omega t + \theta)}. \nonumber 
\end{eqnarray}
The analytical and computational results match well converging to the linearized frequency response (Fig. \ref{fig:figure1}b, c) as $b \rightarrow 0$,
\begin{eqnarray}
\lim_{b\rightarrow0}G(j\omega, b, k_{+0}) &=& \dfrac{A_Tk_-/(k_-+k_{+0})}{\sqrt{\omega^2 + (k_{+0} + k_-)^2}}e^{j(\omega t + \theta)}, \nonumber 
\end{eqnarray}
with the operating point $(k_+ = k_{+0}, A_1 = A_Tk_+/(k_+ + k_-))$.\\
\emph{Parametric Dependence of Approximation}. We note that the phase of the approximation is independent of the forcing amplitude $b$ and coincides with that of linearization.
At frequencies lower than crossover frequency ($\omega_0$), the output is in phase, whereas at higher frequencies the output lags the input by $\pi/2$.
The magnitude, on the other hand, depends on forcing amplitude $b$, in an increasing manner (Eq. (\ref{eq:eqn2})). This is a counterintuitive result as steady-state solution of Eq. (\ref{eq:eqn1}),
\begin{eqnarray}
A_1 = A_T\dfrac{k_+}{k_++k_-}, \nonumber
\end{eqnarray}
is an increasing and then saturating function of $k_+$.
However, $A_{1b}$ itself decreases as $b$ is increased.This is because $A_{10}$ decreases as the forcing amplitude $b$ increases (Eq. (\ref{eq:eqn2})).
Increasing either $k_{+0}$ or $k_-$ increases the crossover frequency ($\omega_0 = k_{+0} + k_-$), making the output phase similar to input phase.
In the following two limits,
\begin{eqnarray}
\omega_0 \ll \omega & \Longrightarrow & \dfrac{A_{1b}}{b} = A_T\dfrac{k_-}{\omega_0\omega}, \nonumber \\
\omega_0 \gg \omega & \Longrightarrow & \dfrac{A_{1b}}{b} = A_T\dfrac{k_-}{\omega_0^2 - b^2/2}, \nonumber
\end{eqnarray}
magnitude decreases with $k_{+0}$. But, in the high frequency limit, magnitude increases with $k_-$, whereas in the low frequency limit, the magnitude first increases and then decreases as $k_-$ is increased.

Finally, $A_T$, the total protein concentration scales the magnitude but does not change the phase. 

This presents an overview of how the describing function approximation depends on system parameters.
We note that these relations are more accurate than when obtained by standard linearization and cannot be achieved through numerical simulation.\\
\emph{Approximation Error}. It is important to quantify the error in any approximation as a measure of its accuracy and to compare different approximation methods.
We computed the relative mean square error,
\begin{eqnarray}
e = \dfrac{1}{A_{1b}}\sqrt{\dfrac{1}{T}\int_0^T(y(t) - \tilde{y}(t))^2dt},
\label{eq:eqn3}
\end{eqnarray} 
\noindent where, $y(t)$ is numerical solution of Eq. (\ref{eq:eqn1}) with sinusoidal forcing and,
\begin{eqnarray}
\tilde{y}(t) = A_{10} + A_{1b}\sin(\omega t + \theta), \nonumber
\end{eqnarray}
which can be found analytically from Eq. (\ref{eq:eqn2}) or can be computed using Eq. (\ref{eq:ComputationalDF}) through computer simulation. A comparison of relative error for different approximations (Fig. \ref{fig:figure1}d) shows that the computational describing function-based approximation matches reasonably well with the analytical approximation especially at lower frequencies and is  better than the approximation obtained by the standard linearization.

These results provide a describing-function based approximation of a simple system as well as a quantification of the error involved.\\

\noindent \emph{Example 2: Biomolecular Covalent Modification System with Switch-like Input-Output Map}

As a second example of a biomolecular signaling system, we investigated a covalent modification scheme that is present in multiple cellular pathways~\cite{c8}. This is similar to the example considered above, with enzyme-catalyzed interconversion reactions.
Depending on parameter regime of operation, the steady-state input-output response may have different sensitivities. The reaction scheme for this system is, 
$$A+E_{1} \begin{array}{c}
a_{1}\\
\rightleftharpoons\\
d_{1}
\end{array}AE_{1}
\begin{array}{c}
k_{1}\\
\rightarrow \\
  ^{}\end{array}A^{*}+E_{1} $$
  
 $$A^{*}+E_{2} \begin{array}{c}
a_{2}\\
\rightleftharpoons\\
d_{2}
\end{array}A^{*}E_{2}
\begin{array}{c}
k_{2}\\
\rightarrow \\
  ^{}\end{array}A+E_{2}. $$ 
The forward and backward conversion between the two forms $A$ and $A^*$ are catalyzed by enzymes $E_1$ and $E_2$, respectively.
As such, these reactions have intermediates $[AE_1]$ and $[A^*E_2]$, giving overall conservation relations as,
\begin{align*}
A_{T}= & [A]+[A^*]+[AE_{1}]+[A^{*}E_{2}],\\
E_{1T}= & [E_{1}]+[AE_{1}],\\
E_{2T}= & [E_{2}]+[A^{*}E_{2}],
\end{align*}
where $A_{T}$ is the total substrate concentration and $E_{1T}$ and $E_{2T}$ are the total enzyme concentrations. 
Using these the mathematical model can be obtained as,
\begin{align}
\begin{split}
\dfrac{d[A]}{dt} ={}& -a_{1}[A](E_{1T}-[AE_1])+d_{1}[AE_{1}]\\ \nonumber
                   &+k_{2}(A_T-[A]-[A^*]-[AE_1]),
\end{split}\\
\begin{split}
\dfrac{d[A^{*}]}{dt} ={}& -a_{2}[A^{*}](E_{2T}-A_T+[A]+[A^*]+[AE_1])\\ \nonumber
                       & +d_{2}(A_T-[A]-[A^*]-[AE_1])+k_{1}[AE_{1}],
\end{split}\\
\dfrac{d[AE_{1}]}{dt} ={}& a_{1}[A](E_{1T}-[AE_1])-(d_{1}+k_{1})[AE_{1}].
\label{eq:eqn4}
\end{align}
\begin{figure}[!htb]
\center
\includegraphics[scale=0.34]{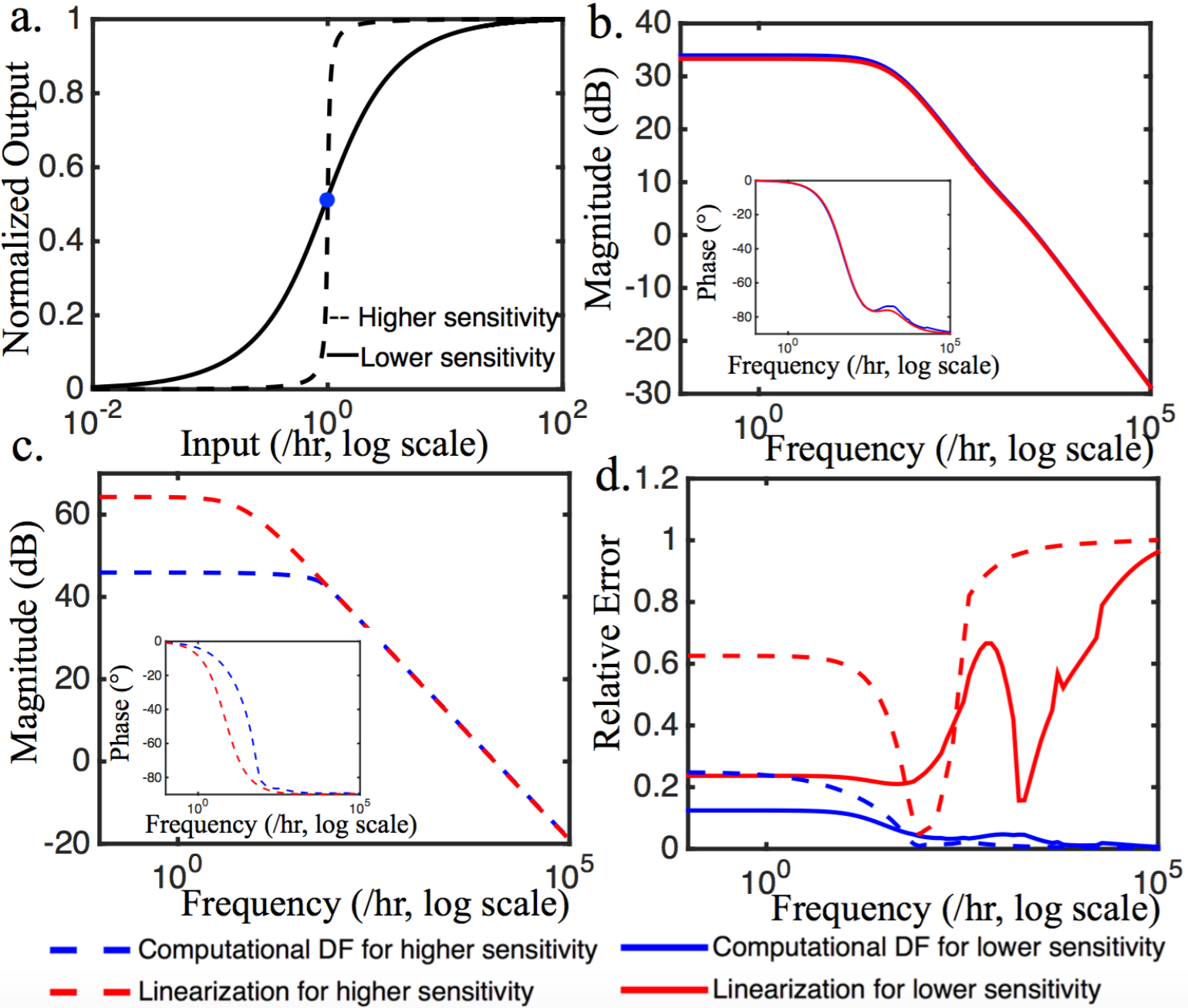}
\caption{Approximation of the biomolecular covalent modification system and error estimates. a. Solid black line represents the steady-state normalized input-output map for parameters $A_{T}=200 nM$, $E_{1T}=E_{2T}=20 nM$, $d_{1}=d_{2}=1 /hr$, $k_{2}=1 /hr$, $k_{10}=k_2$, $b=k_{10}/2$ and $a_{1}=a_{2}= 10^{-2} /hr$, corresponding to a less steep transition. Dashed black line represents the steady-state normalized input-output map which corresponds to a steeper transition, with $a_{1}=a_{2}= 1 /hr$ and other parameters are kept constant. Blue dot represents operating point used for computation of linearization. b. Blue solid line and red solid line represent the magnitude plots of computational describing function-based approximation and linearization approximation of less steep transition respectively, frequency is varied logarithmically in the range $0.1/hr$--$10^4/hr$. Inset shows corresponding phase plots. c. Blue and red dashed lines represent magnitude and phase plots for both approximations in steeper transition case. d. Corresponding error involved in both approximations for both sensitivity regimes.} 
\label{fig:figure2}
\end{figure}
As before, we model $k_1$ as the input and the concentration of $A^*$ as the output.
The input-output response at steady-state can have different sensitivities depending on the parameter regime of operation.
Two such choices with a low sensitivity and a high sensitivity are shown in Fig. \ref{fig:figure2}b, c.
The nonlinearities are distributed through the system and arise due to the principles of mass action.
We aim to approximate the overall input-output response with a describing function-based linearization.
Therefore, while this is similar to a saturation nonlinearity~\cite{c7}, there are inherent dynamics.\\
\noindent \emph{Calculation of Approximation}. The describing function approximation is obtained computationally (Fig. \ref{fig:figure2}b, blue solid line; \ref{fig:figure2}c, blue dashed line), as described previously.
We also computed, for comparison, the standard linearization (Fig. \ref{fig:figure2}b, red solid line; \ref{fig:figure2}c, red dashed line).\\
\emph{Parameter Dependence}. The describing function-based approximations in both regimes have a similar low pass nature.
The DC gain for the higher sensitivity regime (Fig. \ref{fig:figure2}c) is higher than that that of the lower sensitivity regime (Fig. \ref{fig:figure2}b), consistent with the slope of the operating point.
For the low sensitivity regime, the describing function-based approximation is similar to that of the standard linearization.
Interestingly, however, the DC gain in the higher sensitivity regime is lower than that of the corresponding standard linearization, while the bandwidth is larger.
This shows that, for finite inputs, the gain may not be as high as the standard linearization predicts.
Similarly, the bandwidth may not decrease to a large extent.
Therefore, analysis of the describing function-based approximation provides insight into how the system may behave in case of finite inputs. We emphasize that this realization is not possible through numerical simulations and the describing function technique provides a better approximation than standard linearization. \\
\emph{Approximation Error}. Based on the computation, (Fig. \ref{fig:figure2}d) it can be observed that the error in the describing function-based approximation is lower than that obtained from the standard linearization in both parameter regimes considered.\\

\noindent \emph{Example 3: Biomolecular System with Hysteretic Input-Output Map}

\begin{figure}[!htb]
\center
\includegraphics[scale=0.32]{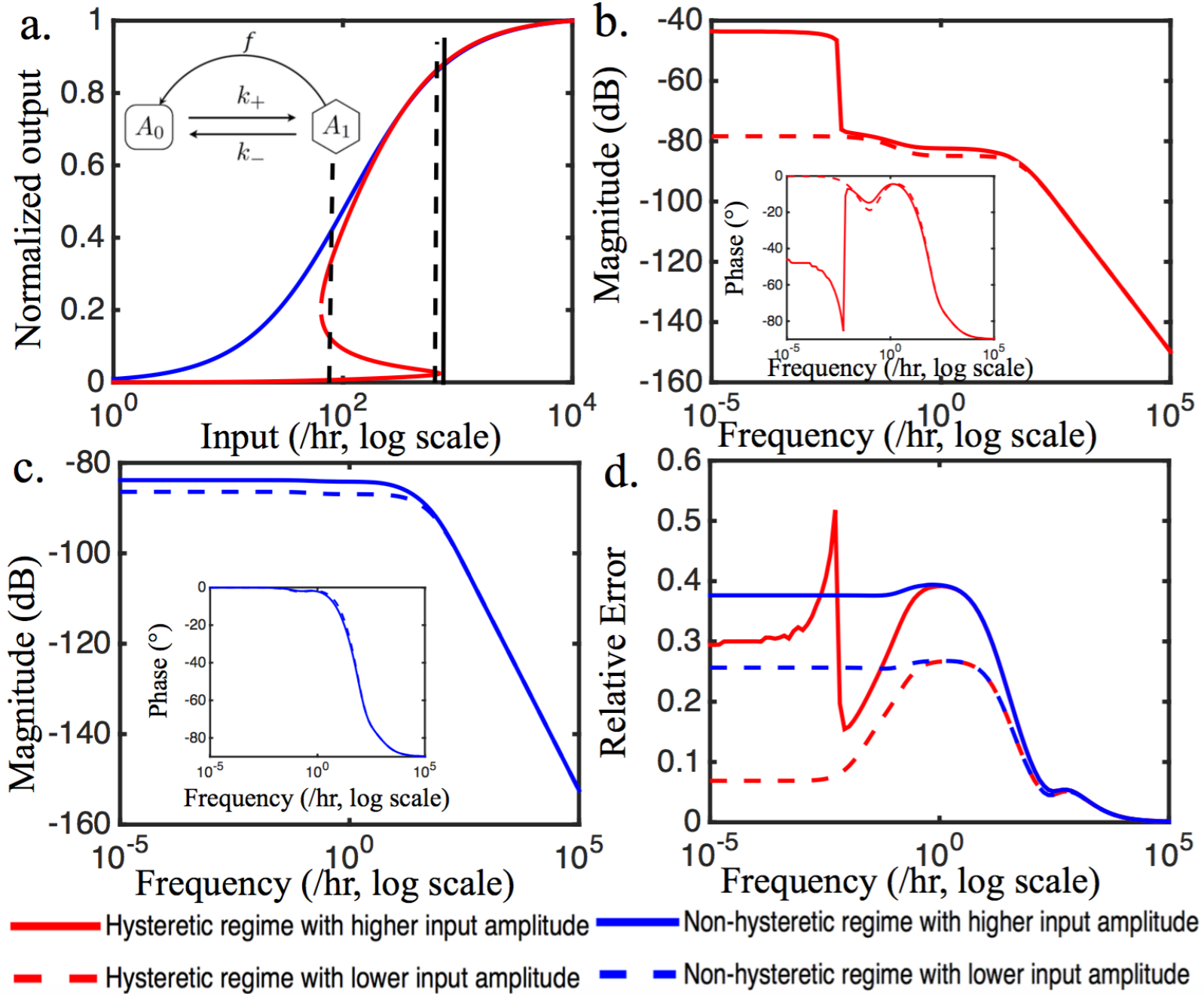}
\caption{Describing function approximation and error estimate for hysteretic system. a. Solid red line represents the steady-state normalized input-output map for parameters $\alpha_0=1/15$, $\alpha=5 nM/hr$, $n=2$, $\gamma=1/hr$, $K=1 nM$, $k_- = 100/hr$, corresponding to a hysteretic response. Solid blue line represents the steady-state normalized input-output map when $\alpha=1$ with other parameters same as above. Solid black vertical line corresponds to the upper limit of the area where describing function approximation is done for higher input amplitude, $k_{+0}=b=375$. Dashed black vertical lines corresponds to the upper and lower limit of the area where describing function approximation is done for lower input amplitude, $k_{+0}=375, b=300$. Schematic of the system is shown in inset. b. Red solid line and dashed line represent the magnitude plot of computational describing function-based approximation for hysteretic response with higher and lower input amplitude respectively and inset shows phase plot, frequency is varied logarithmically in the range $10^{-5}/hr$--$10^5/hr$. 
c. Blue solid and dashed line shows the magnitude and phase plots for non-hysteretic regime with higher and lower input amplitude respectively. d. Corresponding error associated with the approximation in both regime for both higher and lower input amplitude.}
\label{fig:figure3}
\end{figure}

Another example of a biomolecular input-output response, often encountered  in developmental contexts, is hysteresis. One of the simplest ways through which this can be achieved is, through the addition of transcriptional positive feedback in the circuit showed in Example 1. The resulting all-or-none behaviour has been experimentally observed during maturation of Xenopus oocytes \cite{c9,c10}. Here, we consider a simple system to illustrate the hysteretic input-output map (Fig. \ref{fig:figure3}a, inset).
\begin{align}
\begin{split}
\dfrac{d[A_{1}]}{dt} ={}& k_{+}(A_{T}-A_{1})-\gamma A_{1}-k_{-}A_{1},\nonumber
\end{split} \\        
\dfrac{d[A_{T}]}{dt} ={}& f(A_{1})-\gamma A_T, 
\label{eq:eqn5}
\end{align}

\noindent where the feedback is $f(A_1)=\alpha_{0} + \dfrac{\alpha (A_1)^n}{K^n+(A_1)^n}$ and there is the conservation law, $A_T=A_0+A_1$. The addition of the nonlinear ultrasensitive ($n>1$) positive feedback can generate a hysteretic response (Fig. \ref{fig:figure3}a). In the hysteretic regime, there are three steady-states, two of which are stable and one is unstable. As input parameter $k_+$ is varied, a pair of stable-unstable steady-states coalesce causing a monostable high or low steady-state. Our aim is to develop an approximation of the entire input-output map from $k_+$ to $A_1$. While the hysteretic map looks like the classical ones analyzed, the difference here is that there are inherent dynamics in the hysteresis nonlinearity.\\
\emph{Calculation of Approximation}. The describing function approximation is computed by setting the input, $k_{+}$, as a biased sinusoid and taking the first harmonic approximation of the output $A_1$. We analyzed parameter regimes where hysteretic response is present as well as the one in which it is not present (Fig. \ref{fig:figure3}b, c). In fact, due to the presence of multiple steady-states in the hysteretic regime, it is unclear which point to take for the standard linearization. For the describing function-based approximation, however, there is no such ambiguity.

\noindent \emph{Parameter Dependence}. The frequency response shows low pass filter nature in both regimes. We note that the frequency response in the hysteretic regime can have a low frequency part where the magnitude is constant and the phase is decreasing (Fig. \ref{fig:figure3}b, solid red line). This is exactly the characteristic associated with a pure delay (Transfer function $H(j\omega) = e^{-j\omega T} \implies |H(j\omega)| = 1, \angle H(j\omega) = -\omega T$). Therefore, this method of approximating a positive feedback loop adds insight to system analysis by providing a direct map to a pure delay as well as an estimate of the quantitative parameters associated with it. This is consistent with other studies relating positive feedback with a delayed action, for example~\cite{trotta2010delayed}. With lower input amplitude we still observe two humps in the magnitude and phase plot (Fig. \ref{fig:figure3}b), possibly owing to the presence of two time scales --- that of protein production and covalent modification. An attenuated version of this trend is also visible in the case when hysteretic response is absent (Fig. \ref{fig:figure3}c). Although, the higher and lower input amplitude response does not have much difference when there hysteresis is not expected (Fig. \ref{fig:figure3}c).\\
\emph{Approximation Error}. We calculated the error involved (Fig. \ref{fig:figure3}d) in approximation for the hysteretic input-output map for different regimes and different input amplitude chosen in the input-output plot of Fig. \ref{fig:figure3}a. The error is higher for higher input amplitude in the hysteretic regime owing to the presence of strong nonlinearity. 

\section{Analytical error bound}
\label{sec3}
\noindent As with any approximation method, it is important to estimate the associated error. 
As the describing function technique essentially approximates a dynamical system by the first harmonic of its output, when a sinusoidal input is applied, we develop an error bound using methods associated with Fourier series. The Fourier series of a periodic signal is,\\
$$f(t) = a_0 + \sum_{n=1}^{\infty} ( a_{k}\cos n\omega_0 t+b_{k}\sin n\omega_0 t ).$$\\
This can also be represented in exponential form,\\
$$f(t)=\sum_{n=-\infty}^{\infty} C_{n}e^{ jn\omega_{0}t}, \quad \omega_{0}=\dfrac{2\pi}{T},$$\\where,
\begin{align}
C_{n}=\dfrac{1}{T}\int_{-T/2}^{T/2} f(t)e^{-jn\omega_{0}t}dt.
\label{eq:eqn6}
\end{align}
From the convergence of Fourier series, we know that signals that are continuous or have a finite and bounded discontinuity over a period satisfy~\cite{oppenheim}, $$\lim_{m\to\infty}\|f-f_m\|=0,$$ where $f_{m}(t)=\sum_{n=-m}^{m} C_{n}e^{ jn\omega_{0}t}$, is the Fourier series taken upto $m^{th}$ harmonic.
For the describing function case, we need to estimate  $$\epsilon_1 =  \|f-f_1\|=\int_{-\pi}^{\pi}|f-f_{1}|^2 dt,$$ which is basically a scaled version of the error defined in Eq. (\ref{eq:eqn3}). For Fourier series an upper bound of this error can be obtained using the principle of bounded variation~\cite{fourierbound}.

\begin{definition}[\cite{variationthesis}]
Let $f:[a,b] \rightarrow \mathbb{R}$ be a function and $M=\{x_0 , x_1 ,.., x_n \}$ be a partition of $[a,b].$ We define the total variation of $f$ over the $[a,b]$ as, $$V_a^b (f)=\sup_{M} \left\lbrace \sum_{k=0}^{n-1}|f(x_{k+1})-f(x_k)|\right\rbrace,$$ where the supremum is taken over all partitions of $f.$ The function is said to have bounded variation if $V_a^b (f)$ is finite in $[a,b]$ and we write $f \in V[a,b].$
\end{definition}
 
The following properties are useful in calculating the total variation~\cite{variationthesis},\\
1. If $f:[a,b] \rightarrow \mathbb{R}$ is monotone in $[a,b]$, then $f \in V[a,b]$ and $V_a^b (f)=|f(b)-f(a)|.$\\
2. Let $f:[a,b] \rightarrow \mathbb{R}$ be of bounded variation in $[a,b].$ Then $cf \in V[a,b]$ for any $c \in \mathbb{R}$ and $V_a^b (cf)=|c|V_a^b (f).$\\
3. Let $f:[a,b] \rightarrow \mathbb{R}$ and $c$ is an arbitrary point in $[a,b]$. Then $f \in V[a,b]$ if and only if $f \in V[a,c]$ and $f \in V[c,b].$ Furthermore, $V_a^b (f)=V_a^c (f)+V_c^b (f).$\\
\begin{theorem}[\cite{fourierbound}]
If $f(t)$ is periodic with frequency $\omega_0$ and the total variation over one period is bounded by $V$, then the mean square error in approximating upto $m^{th}$ harmonic is, $$\epsilon_{m}^{2} \leq \dfrac{V^2}{\pi \omega_{0}m}.$$
\end{theorem}

We use this result to develop an error bound for the describing function approximation.
We apply this to classically known static nonlinearities and then illustrate how this can be applied to the dynamic nonlinearities such as, the biomolecular systems discussed in Section \ref{sec2}.\\

\noindent \emph{Example 4: Static Saturation Nonlinearity}

Consider a saturation nonlinearity with $a$ and $k$ denoting the range and slope of saturation. To obtain the describing function approximation we use a sinusoidal input $A\sin \omega t$ to the nonlinearity and calculate the first harmonic (\cite{c7}) of the output.
The output of the nonlinearity is described as,
\[
    f(t)= 
\begin{cases}
    kA\sin \omega t,& \text{if } 0\leq \omega t \leq \delta\\
    ka,              & \text{if } \delta < \omega t \leq \dfrac{\pi}{2}, \text{where, } \delta = \sin^{-1} (\dfrac{a}{A})
\end{cases}
\]
Using the properties of bounded variation,
\begin{align*}
& V_0^\delta (f)=f(\delta)-f(0)=kA\sin \delta -0 = ka,\\
& V_\delta^{\pi -\delta}=0,\\
& V_{\pi - \delta}^{\pi}= ka.
 \Rightarrow V_0^{2\pi}=4ka.
\end{align*}

\begin{figure}[!htb]
\center
\hspace*{-0.1cm}\includegraphics[scale=0.38]{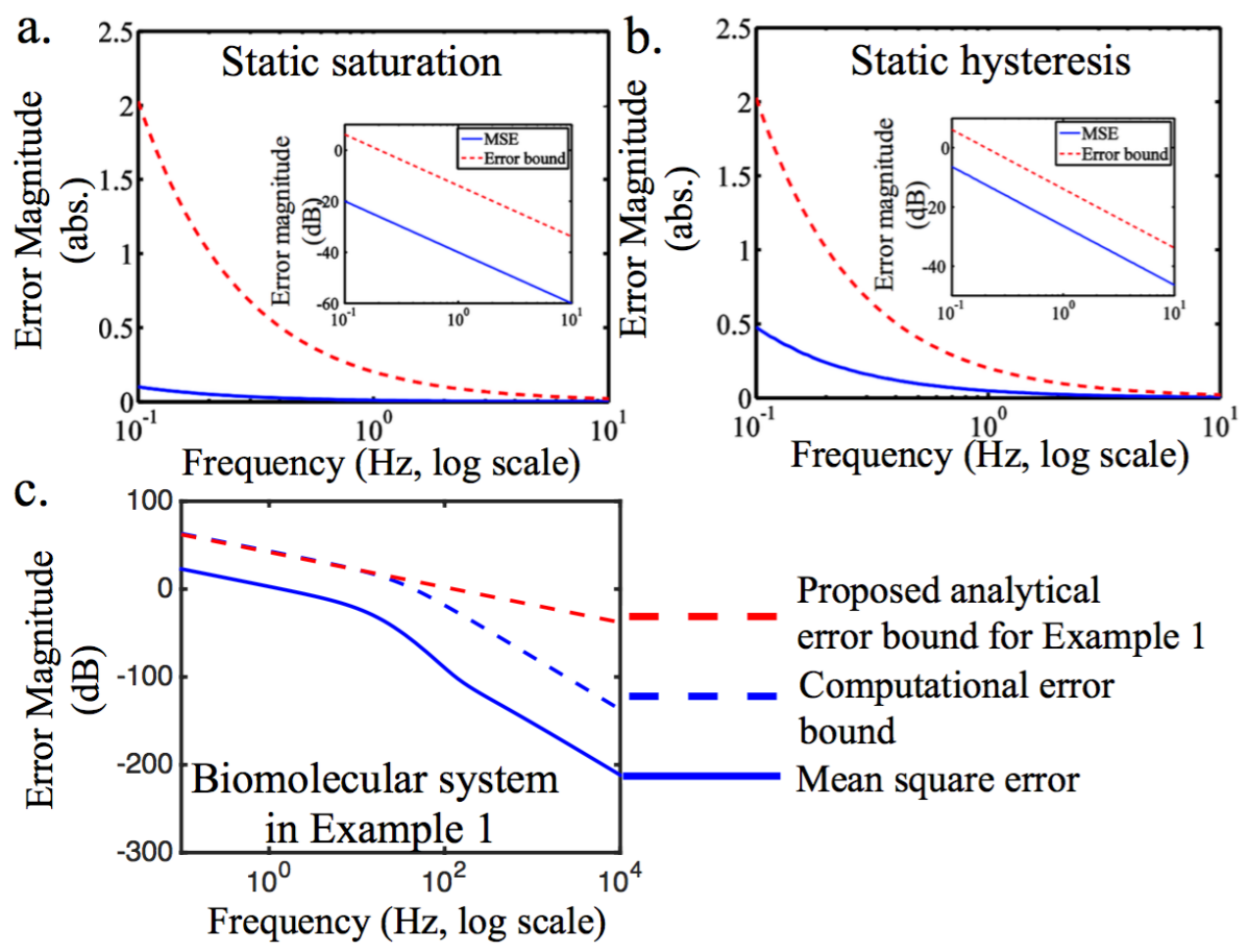}
\caption{Error in describing function approximation and corresponding error bounds. a. Blue solid line and red dotted line represents mean squared error in describing function approximation and corresponding error bound respectively for static saturation, when $\omega$ is varied. The same is shown in dB scale in the inset. b. Similar plots for static hysteresis case. c. For the input-output map in Example 1, solid blue line represents the mean square error in the describing function approximation. Red dashed line represents the theoretical error bound when total variation is taken as $A_{T}/2$. Dashed blue line represents the bound when total variation is calculated computationally.}
\label{figure:fig5}
\end{figure}

\noindent In fact, for any real periodic signal, the total variation over a period is bounded and is four times of the amplitude of the signal.
Therefore,\\
\begin{center}
$\epsilon_{1}^2 = \|f-f_1\|_{2}^2 \leq \frac{(4ka)^2}{\pi \omega (1)}$,
\end{center}
 for approximation upto first harmonic.
We find that the calculated mean squared error in the first harmonic approximation is bounded by the theoretical error bound for static saturation nonlinearity as frequency is varied (Fig. \ref{figure:fig5}a).\\
\noindent \emph{Example 5: Static Hysteresis Nonlinearity}

A relay exhibiting hysteresis is another commonly encountered static nonlinearity (\cite{c7}). Parameters $\delta$ and $D$ denotes the hysteretic angle  and the range of saturation respectively. It works like a simple relay with a phase shift of $\delta$.
The error bound in this case is, $\epsilon_{1}^2 = \|f-f_1\|_{2}^2 \leq \dfrac{(4D)^2}{\pi \omega (1)}.$ Fig. \ref{figure:fig5}b  shows the mean square error between actual output and the first harmonic approximation which is bounded by theoretical error bound for all  $\omega$.
It can be noted that the error bound for both cases are similar if the maximum magnitude of the output is set to be same in both cases, i.e. $ka=D.$ However, it is observed that the maximum mean squared error for the relay with hysteresis is more than that of static saturation nonlinearity. The maximum mean squared error increases for saturation case if the slope of saturation increases and tends to the error of relay with hysteresis when $k$ approaches $\infty$.\\

\noindent \emph{Example 1 (continued)}

The nonlinearities encountered in biomolecular systems often have a dynamic character, as analyzed in previous section. To find an error bound for these kind of nonlinearities, we first need to determine the total variation. We illustrate this for the first example, where we bound the maximum variation for the output $A_1$ to be  $A_T$, the total concentration. Further, the steady-state input-output response for static saturation case does not take negative values. Comparing these, we find that the total variation of periodic output in this case is, $V=A_{T}/2$ and the error bound is $V^{2}/\pi\omega$. Computationally, we find the total variation from the time trajectory of simulated output. The theoretical as well as computational error bounds for describing function approximation are shown in Fig. \ref{figure:fig5}c. Both of these bound the mean square error of describing function approximation.

These results provide an error bound for describing function approximation using the concept of total variation and a general way, using the maximum allowable concentration of the output species, obtain these error bounds for biomolecular systems.

\section{Error estimates in limit cycle analysis}
\label{sec4}

Finally, we investigate the use of the error analysis developed above in the analysis of limit cycles, a classically important application of the describing function technique. Although this technique is widely used in control engineering for limit cycle prediction, a treatment of errors involved is relatively less common~\cite{dfbound}. Here, we aim to incorporate the simple error estimate developed above to augment the standard analysis.\\

\noindent \emph{Example 6: Van der Pol Oscillator} 

We begin with the Van der Pol oscillation, a benchmark nonlinear oscillator and commonly used to illustrate the describing function-based limit cycle analysis~\cite{c7}. This oscillator was first analyzed in the context of vacuum tube triode circuits and termed as `relaxation oscillation' by Van der Pol~\cite{van_der_pol}.  This class of oscillators has multiple applications in physical as well as biological sciences, such as in the Fitzhugh-Nagumo oscillator~\cite{fitzhugh, nagumo}. The dynamical equations of Van der Pol oscillator are,
\begin{equation}
\ddot{x}+\mu(x^{2}-1)\dot{x}+x=0
\end{equation}
\begin{figure}[!htb]
\center
\hspace*{-0.2cm}\includegraphics[scale=0.35]{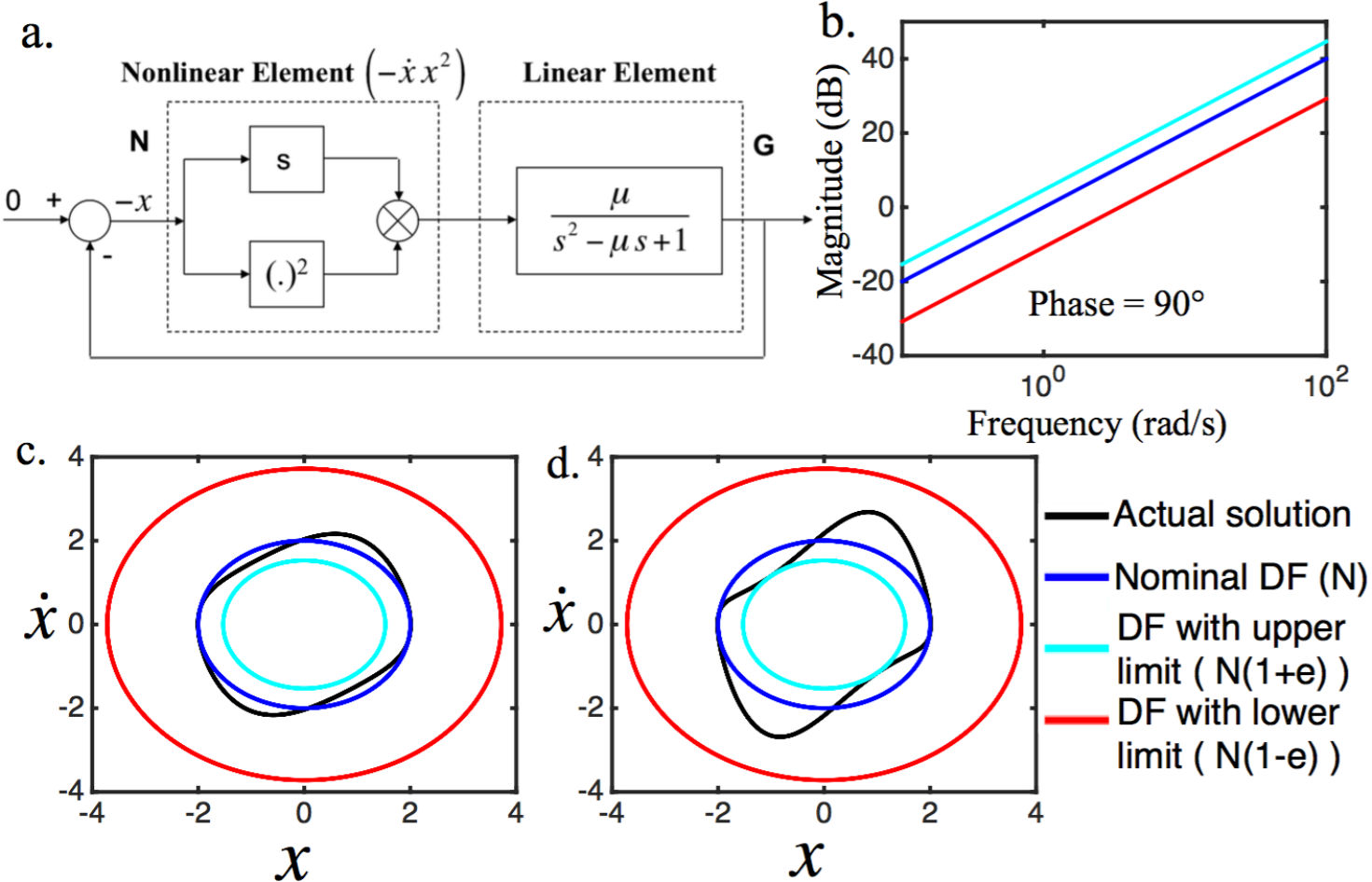}
\caption{Limit cycle analysis of Van der Pol oscillator using describing function technique. a. Block diagram of Van der Pol oscillator. b. Solid blue line represents nominal describing function approximation whereas  upper and lower limit for the approximation is represented by cyan and red line respectively. c. Black line represents the actual solution for Van der Pol oscillator. Blue, red and cyan line corresponds to the solution considering nominal, upper and lower limit of describing function approximation for $\mu = 0.4$. d. The same is repeated for  $\mu =1$.}
\label{vdpdf}
\end{figure}
This can be separated in a linear and a nonlinear part as shown in Fig. \ref{vdpdf}a.\\
\noindent \emph{Describing Function Approximation with Error Estimate}. We computed the describing function of the nonlinear part (Fig. \ref{vdpdf}b). This is denoted $N$. We also computed the error ($e$) in this approximation to get an uncertainty model $N(1 \pm e)$, where the $+$ and $-$ signs give, respectively, the upper and lower limits.
Next, we used the condition for sustained oscillation, $1+NG=0$, where $G$ is the transfer function of the linear part of the loop. We did this for all three cases \--- nominal ($N$), upper limit ($N = N(1 + e)$), lower limit ($N = N(1 - e)$) \--- to get the corresponding values of the oscillation amplitude ($A$).The Nyquist plot corresponding to the nominal describing function approximation touches the $-1+j0$ point at $\omega = 1$ rad/s for $A = 2$.
This is consistent with the classical describing function analysis that predicts an oscillation frequency of $\omega = 1$ rad/ s and an oscillation amplitude of $2$.  
Incorporating the error information ($N(1 \pm e)$) in the condition of oscillation, we obtain two other values of oscillation amplitude, $A = 1.529$ and $A = 3.718$.
These suggests that the actual oscillations might lie in this uncertainty band.

To check this, we computed the actual solution of the Van der Pol oscillator along with the solution obtained from the nominal as well as the error analysis for $\mu  = 0.4$ and $\mu = 1$ (Fig, \ref{vdpdf}c, d).
At lower values of $\mu$, the actual solution matches the nominal describing function prediction, although the extent of deviation increases as $\mu$ increases.
In fact, the upper and lower amplitudes provide bounds for the variation of the actual amplitude over a period for these parameter choices.
Therefore, these error estimates can be used in predicting an uncertainty band of oscillation amplitudes, providing a larger range of parameters where the describing function approximation can help in the limit cycle analysis.

We note that \cite{dfbound} have also performed a general error calculation of limit cycle, which in this case yields error bounds on amplitude and frequency for a larger range of parameter values. Their approach is based on analyzing approximation error and its propagation in the entire closed loop. Here, we focus on the bounds simply obtained due to the uncertainty representation of the nonlinearity.

\noindent \emph{Example 7: Biomolecular Ring Oscillator}

\begin{figure}[!htb]
\center
 \includegraphics[scale=0.3]{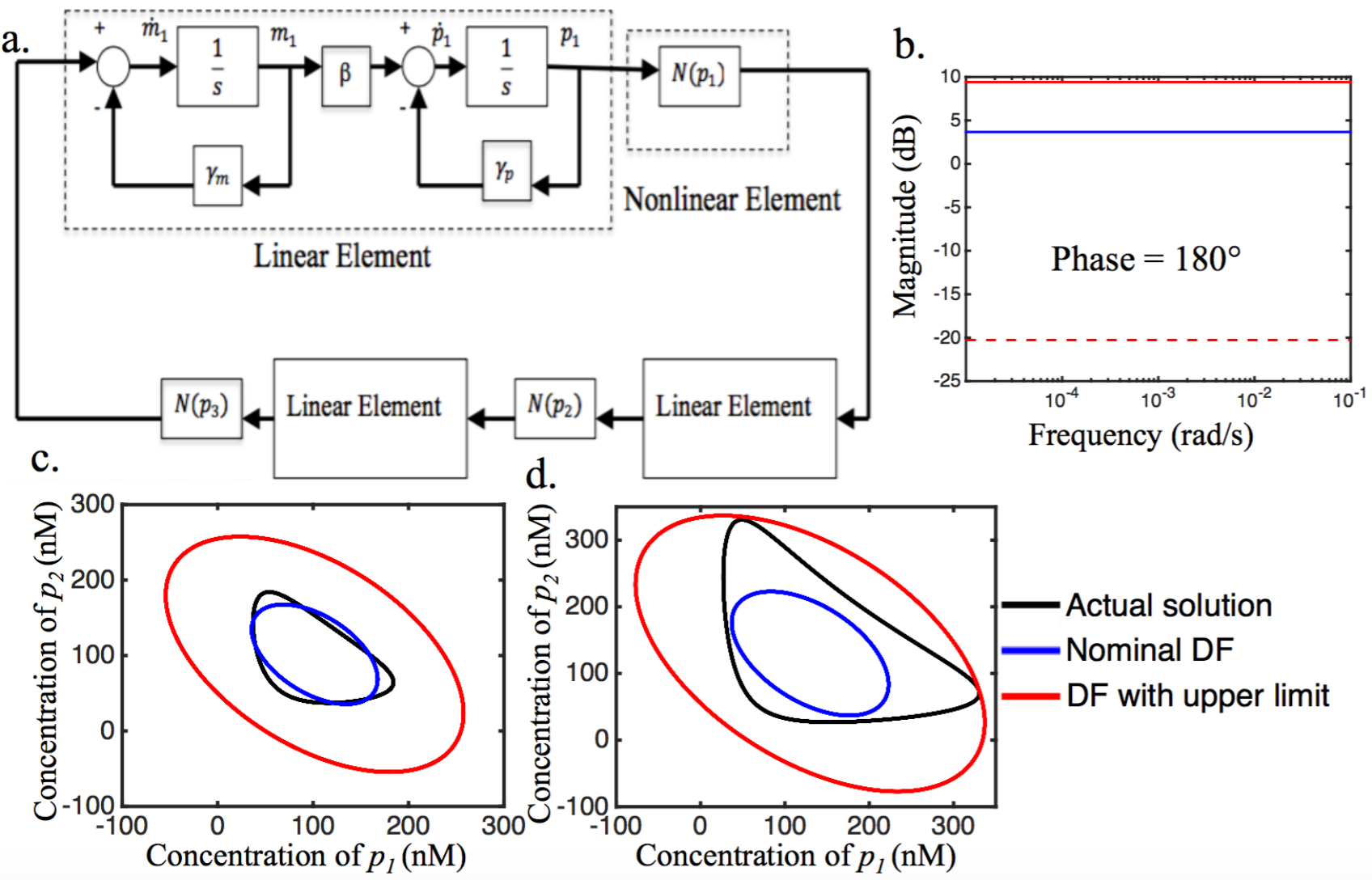}
\caption{Limit cycle analysis of repressilator using describing function technique. a. Block diagram of repressilator. b. Solid blue line represents nominal describing function approximation whereas  upper and lower limit for the approximation is represented by solid red and dashed red line respectively for parameter values $\alpha = 0.4995$ nM/s, $\alpha_{0}=5 \times 10^{-4}$ nM/s, $\gamma_{m}=\log 2 /120$ $s^{-1}$, $\gamma_{p}=\log 2/600$ $s^{-1}$, $\beta=\log 2/6$ $s^{-1}$, $K=40$ nM, $n=2$. c. Black, blue and red lines represent the actual solution, nominal describing function solution and describing function solution considering the upper error limit for repressilator when $\gamma_{p}=20 \log 2/600$ and other parameters are same as b. d. The same is repeated for $\gamma_{p}=14 \log 2/600$.}
\label{repressdf}
\end{figure}

The repressilator is a biomolecular oscillator consisting of three proteins that repress each other in a ring~\cite{repress}. It is a benchmark oscillator in the field of synthetic biology --- design using biomolecular substrates --- for its demonstration of engineering rich dynamic behaviour inside cells.
The dynamical equations for the repressilator are,

\begin{align}
\begin{split}
\dfrac{dm_{1}}{dt} ={}& -\gamma_{m} m_{1}+\alpha_{0}+\dfrac{\alpha}{1+(p_{3}/K)^n},\nonumber
\end{split} \\ 
\begin{split}       
\dfrac{dp_{1}}{dt} ={}& -\gamma_{p}p_{1}+\beta m_{1},\nonumber
\end{split}\\
\begin{split}
\dfrac{dm_{2}}{dt} ={}& -\gamma_{m} m_{2}+\alpha_{0}+\dfrac{\alpha}{1+(p_{1}/K)^n},\nonumber
\end{split} \\ 
\begin{split}       
\dfrac{dp_{2}}{dt} ={}& -\gamma_{p}p_{2}+\beta m_{2},\nonumber
\end{split}\\
\begin{split}
\dfrac{dm_{3}}{dt} ={}& -\gamma_{m} m_{3}+\alpha_{0}+\dfrac{\alpha}{1+(p_{2}/K)^n},\nonumber
\end{split} \\      
\dfrac{dp_{3}}{dt} ={}& -\gamma_{p}p_{3}+\beta m_{3}.
\label{eq:eqn8}
\end{align}
The form of these equations and the nominal parameters are taken from~\cite{repress}.
Briefly, $m_i$ and $p_i$ ($i = 1, 2 ,3$) represent the concentrations of mRNA and protein, respectively.
For simplicity the production and degradation parameters are assumed to be same for all three mRNA - protein pairs.
$\gamma_p$ and $\gamma_m$ represent the degradation rate constants for protein and mRNA, respectively.
$\beta$ is the translation rate constant.
The repression function for the $i^{th}$ mRNA $\alpha_0 + \alpha/ (1+p_j/K)^n)$, models the repression of $m_i$ as $ p_j$ increases ($j = 2$ for $i =1$, $j = 3$, for $i = 2$, $j= 1$ for $i =3$). This is a special case of a general cyclic gene regulatory network which may give rise to oscillations and the existence criteria and general oscillation profiles are given in~\cite{hori} using harmonic balance analysis. 
To apply the describing function technique and associated error analysis, we separate the system to linear and nonlinear parts (Fig. \ref{repressdf}a).
There are three transcriptional modules, each of which has one linear and one nonlinear part which is $F(p) =  \alpha_0 + \dfrac{\alpha}{(1+p_j/K)^n}$ ($j = 1,2,3$).
The gain of the describing function approximation for a single nonlinear element is denoted as N.
The input to each nonlinear part is concentration of a protein and thus it is set to a biased sinusoidal input, $p_i = a + b \sin(\omega t)$. 
Unlike the previous case, the input has a bias part.
This is because the concentrations in biomolecular systems concentrations can not be negative ($b \leq a$).

\noindent \emph{Calculation of Frequency and Describing Function}\\
From the condition of sustained oscillation,
\begin{equation}
1+N^3G^3(s)=0, 
\end{equation}
where $G(s)=\dfrac{1}{(1+T_{\gamma_{m}s})(1+T_{\gamma_{p}s})}$ is the linear element in one transcriptional module. From this we get one magnitude and one angle condition,
\begin{eqnarray}
|(1+T_{\gamma_{m}}s)(1+T_{\gamma_{p}}s)| &=N \nonumber \\
3 \angle [(1+T_{\gamma_{m}}s)(1+T_{\gamma_{p}}s)] &= \pi ,
\end{eqnarray}
where, $T_{\gamma_{m}}=\dfrac{1}{\gamma_{m}}$ and $T_{\gamma_{p}}=\dfrac{1}{\gamma_{p}}$.
This yields,
\begin{eqnarray}
\omega &=\dfrac{-(T_{\gamma_{m}}+T_{\gamma_{p}})\pm \sqrt{(T_{\gamma_{m}}+T_{\gamma_{p}})^2+4T_{\gamma_{m}}T_{\gamma_{p}}\tan^{2}\dfrac{\pi}{3}}}{2T_{\gamma_{m}}T_{\gamma_{p}}\tan\dfrac{\pi}{3}},
\label{eq:freq}
\end{eqnarray}
and, 
\begin{eqnarray}
N &= \sqrt{(1-T_{\gamma_{m}}T_{\gamma_{p}}\omega^{2})^2+\omega^{2}(T_{\gamma_{m}}+T_{\gamma_{p}})^2}
\label{eq:gain}
\end{eqnarray}

\noindent \emph{Calculation of the Input Bias}

The simulation suggests that there is no unique input bias and amplitude pair (a,b) which gives rise to sustained oscillation. So, we need to fix the input bias. Considering the first transcription module in Eq. (\ref{eq:eqn8}), the nonlinear part can be approximated to, $F(p_3)=N_{0}-Np_{3}$. $N_{0}$ is the bias term in the approximation and the negative sign suggests $180^0$ phase for describing function approximation (Fig. \ref{repressdf}b). $N$ is calculated from Eq. (\ref{eq:gain}). Now, putting $p_3=a+b\sin(\omega t+\delta)$ and $m_{1}=a_{m}+b_{m}\sin (\omega t+\phi)$, we get,

\begin{dmath*}
\omega b_{m}\cos(\omega t+\phi) = N_{0}-N(a+b\sin(\omega t+\delta))-\gamma_{m}(a_{m}+b_{m}\sin(\omega t+\phi)), \nonumber 
\end{dmath*}
and,
\begin{dmath*}
\omega b\cos \omega t = \beta(a_{m}+b_{m} \sin(\omega t+\phi))-\gamma_{p}(a+b\sin \omega t) \nonumber
\end{dmath*}

Equating the constant terms we get, 
\begin{equation}
N_{0}=(N+\dfrac{\gamma_{m}\gamma_{p}}{\beta})a,
\label{eq:bias}
\end{equation}
which gives an equation of straight line where every point corresponds to the condition of sustained oscillation. Now from simulation also we can compute the value of $N_{0}$ when input bias, $a$ is varied which produces a family of curves with varying dependence of input bias, $a$ on input amplitude $b$. The intersection points correspond to the value of input bias. The simulation results are closest to the intersection points found by this method for $a=b$, when $n=2$. 

\noindent \emph{Describing Function Approximation with Error Estimate}\\
The magnitude of the describing function approximation with the upper and lower error limits is shown in Fig. \ref{repressdf}b.
We find the frequency of oscillation from Eq. (\ref{eq:freq}) and this matches with simulation result. 
The describing function gain is calculated from Eq. (\ref{eq:gain}) and corresponding input bias is calculated from Eq. \ref{eq:bias}. 
From the sustained oscillation condition, we get the nominal amplitude of oscillation.
The upper limit of the error gives an upper limit of amplitude. 
The lower limit of error does not give a clear result as the error magnitude is larger than the nominal value.
 As we have taken same production and degradation constants for all three transcription module, we assume that the phase difference between each protein concentration profile is $2\pi/3$. The actual simulation of the repressilator system is compared with the oscillation profiles obtained from this analysis in Fig. \ref{repressdf}c, d for different parameter values. We conclude that, like in the case of the Van der Pol oscillator, the use of the describing function technique with the simple error model adds to the limit cycle analysis by providing an estimate of the variation in the oscillation amplitude.

 \section{Discussion}
Describing functions can be used to approximate a nonlinear input-output map with its linearization.
Here, we have adapted this method for investigations of biomolecular systems and presented the following three results.
First, we used this technique to approximate representative input-output responses, both saturating and hysteretic, and mapped the dependence of the approximation on system parameters.
Second, we estimated the approximation error, which was smaller than that of a standard linearization, and theoretically developed a way to obtain an upper bound of this error.
Third, we used the computed error estimate to augment the standard limit cycle prediction by providing a simple way to estimate range of oscillation amplitudes possible.
These results should help to augment a framework for approximating biomolecular signaling systems with linearized versions.

It is interesting to note the additional insight that can be obtained by contrasting the describing function-based linearization with the standard linearization.
For example, in the high sensitivity regime of the enzymatic signaling system (Fig. \ref{fig:figure2}c), standard linearization at the operating point would point to a dynamic response which is very slow and has a high gain.
However, an analysis using a finite amplitude input, as in the describing function approximation here, adds insight to the fact that the slow response is present only if the amplitude is infinitesimal.
For more realistic inputs, where the amplitude may be finite, the response is not significantly slower than in the low sensitivity regime (Fig. \ref{fig:figure2}b), although the gain is still higher.
 This provides a holistic understanding of this signaling system. In case of positive feedback systems, describing function approximation gives an account of the delay (Fig. \ref{fig:figure3}b) associated with it in the hysteretic regime. In presence of multiple steady states, when standard linearization is not well-defined, describing function approximation gives a quantitative measure of this delay and can further aid in the design of biomolecular oscillators based on this.

An important task for the future is to obtain describing function approximations for different systems, both for building blocks such as analyzed over here, as well as for analysis of larger systems obtained by combining such smaller systems. Further, it may be useful to compare these representations with relatively recent experimental data on the response of such systems to periodic input~\cite{c11, freq_osmo_adaptation, fluoro_microscopy,bennett2008metabolic}. For example, we have computed describing function approximation for the input-output map of the Galactose metabolism pathway in \emph{Saccharomyces cerevisiae} using a model described in~\cite{bennett2008metabolic} consisting of eighteen differential equations. This is reported in Supplementary material S2. We find that, for this case also describing function serves as a better approximation than standard linearization and can aid the analysis of frequency response data obtained experimentally. It may also be useful to analyse such complex networks by dividing them in multiple stages and replacing each with their linear approximations. Such a cascade network where each stage is a two component signaling system is analysed in Supplementary material S3. We have shown how each stage can be approximated with a describing function-based approximation and the overall approximation has lower error than standard linearization. Similar approach can be adopted for larger biomolecular systems with the aim of enhancing ease of analysis. Finally, the use of random inputs to obtain such describing function approximations may help in treating biomolecular noise that can be potentially important~\cite{noise}.
 
A mathematical framework can help to understand the behaviour of large complex systems and as a tool for design.
Additionally, an appreciation of the limitations of the framework undeniably aids these aims.
Here, we have adapted the method of describing functions for approximating biomolecular systems and estimated the corresponding error.
\begin{figure}[!htb]
\center
\includegraphics[scale=0.29]{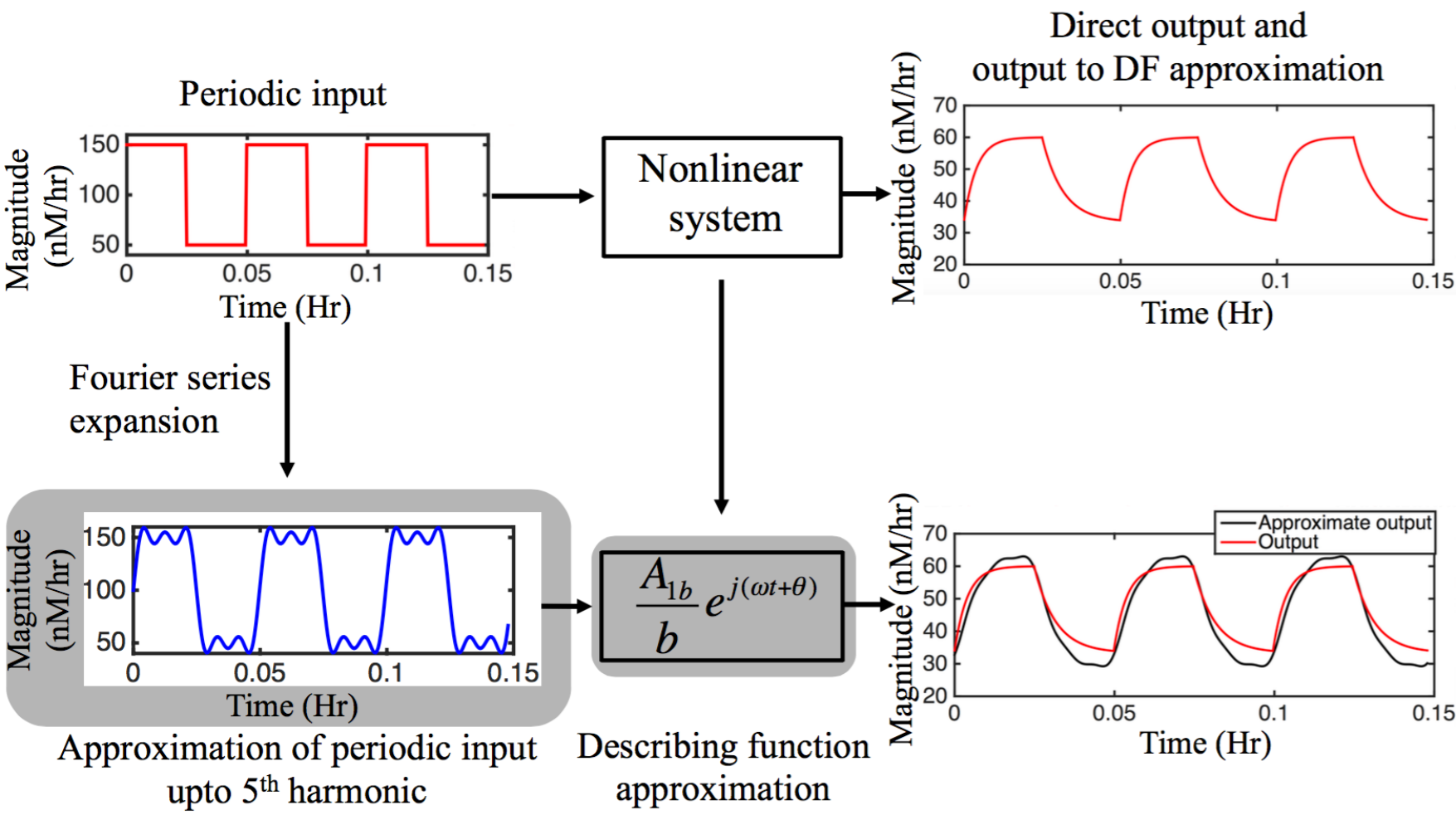}
\caption{Application of describing function technique in estimating output to square wave input for the system in Fig. \ref{fig:figure1}a for $\omega/2\pi=20.0923$ cycles/hour. Square wave input is approximated upto $5$th harmonic and each harmonic is treated as a sinusoidal input for the use of describing function. Output to each harmonic summed together to get the approximate output and the results are compared.}
\label{fig:figure4}
\end{figure}
These can be also used, for example, in estimating the output response to other periodic inputs, such as square waves (Fig. \ref{fig:figure4}).
To illustrate, we compare the output to a square wave to the one obtained by decomposing the square wave into three (fundamental, third and fifth harmonics) Fourier components and then adding the output response of these components as predicted by the describing function.
The output achieved through these two approaches is reasonably similar.
Finally, it should aid in developing a framework for analysis and design of larger, more complex biomolecular systems through systematic interconnections of smaller components.

\section*{Acknowledgement}
Research supported partially by Science and Engineering Research Board grant SB/FTP/ETA-0152/2013.

\bibliographystyle{unsrt}
\bibliography{ref}

\end{document}